\documentclass{aa}
\usepackage{txfonts}
\usepackage{graphicx}
\usepackage{natbib}

\newcommand{\kms}{\,$\mathrm{km\,s^{-1}}$}
\newcommand{\mlp}{\ensuremath{\alpha_{\mathrm{MLT}}}}

\begin{document}

\title{N in turn-off stars of NGC~6397 and NGC~6752
\thanks{Based on observations collected at the ESO VLT, Paranal
Observatory, Chile , program 075.D-0807(A) }}
\subtitle{}
\author{L. Pasquini\inst{1}, A. Ecuvillon \inst{2},
P. Bonifacio\inst{3,4,5}, B. Wolff  \inst{1}}

\offprints{lpasquin@eso.org}

\institute{European Southern Observatory, Garching bei M\"unchen,
Germany
\and
Institut d'Astrophysique, Paris
\and
CIFIST Marie Curie Excellence Team
\and
GEPI, Observatoire de Paris, CNRS, Universit\`e Paris Diderot ; Place Jules Janssen 92190
Meudon, France
\and
Istituto Nazionale di Astrofisica - Osservatorio Astronomico di
Trieste, Via Tiepolo 11, I-34131 Trieste, Italy
}


\titlerunning{Nitrogen in TO stars of Globular Clusters}
\date{Submitted: Accepted}
\abstract
{}
{Our goal is to measure the nitrogen  abundance in 5 Turn Off
(TO) stars of the two Globular Clusters NGC 6397 and NGC 6752, 
and to compare the cluster abundances with those of field stars
of comparable metallicity.
In addition we wish
to investigate the   variations of nitrogen
abundance and its  connections to variations
in the abundances of other light elements.
  }
{We determine the nitrogen abundance from
the compact band head of the
(0-0) vibrational band of the
NH $A^3\Pi-X^3\Sigma^-$ system
at 3360 \AA, using spectra  of resolution R$\sim 45000$ obtained
with the UVES spectrograph on the ESO Kueyen-VLT 8.2m telescope,
analysed with spectrum synthesis based on
plane parallel LTE model atmospheres. We apply the same method previously
used on field stars, to allow a direct comparison of the results. }
{Nitrogen is found to have the same abundance in two of the NGC 6397 stars,
in spite of a  difference  of one order of magnitude in oxygen abundance
between  them. In a third star of the same cluster, the value is  slightly lower, but
compatible with the other two, within the uncertainties.
All the stars in NGC 6397 are N-rich with respect to  field objects
of similar metallicity.
The two stars in NGC 6752  show a difference in 
nitrogen abundance by over one order of
magnitude. The same stars differ
in the abundances of other elements such as 
Na, O and Li,  however   only by a factor $\sim$3-4.}
{NGC 6397 and NGC 6752 are, at present, the only two globular clusters 
in which N has been measured in TO stars in a way consistent
with similar measurements in  field stars. 
The behaviour of N is  different in the two clusters:
no variation is observed NGC 6397, while a large variation
is observed in NGC 6752. This is consistent with a picture in which
the stars in NGC 6752 have been formed by a mixture of ``pristine'' material
and material which has been processed by an early generation of stars,
referred to as ``polluters''.
The N abundances here reported will help to constrain the properties of
the polluters.
In the case of NGC 6397  a simple pollution history is probably
not viable, since the observed variations in O abundances
are not accompanied by corresponding variations in N or Li.
}

\authorrunning{L. Pasquini et al.}
\titlerunning{N in TO stars of globular clusters}
\maketitle


\section{Introduction}

In spite of the fact that nitrogen is one of the  most abundant elements in the
cosmos, only recently it has been
possible to produce  large samples of N abundances in metal poor stars,
enabling us  to
quantitatively study its evolution in the Galaxy (see e.g. Israelian et
al. 2004, and references therein).
Nitrogen tracks iron ([N/Fe]$\sim 0$) down
to a metallicity of $\sim -1.0$. The evolution at lower
metallicity is less clear; while many stars seem to indicate
that N continues to track iron, a few field stars
are found to be nitrogen-rich (the prototypes
of the class
being HD 74000 and HD 160617 \citealt{BN82}),
while others seem to show [N/Fe] ratios below solar.
In their investigation of the chemical composition
in the lowest metallicity stars \citet{spite05}
suggested that a possible interpretation of the
data in the metallicity range
$\rm -4.0 \le [Fe/H]\le -2.5$ is a bimodal
distribution of nitrogen abundances, the majority
of stars at [N/Fe]$\sim +0.4$ and the rest at
[N/Fe]$\sim -1.0$.
Given the complexity of  its nucleosynthesis,
iron is not the best reference element to study
chemical evolution, as discussed in \citet{cayrel04},
and an $\alpha$-chain element such as O or Mg
is preferable.
The [N/$\alpha$] ratio begins to decline
as one moves from solar to lower metallicities
(see e.g. figure 17 of \citealt{spite05}) and at
low metallicities it reaches a constant
value, a {\em plateau}
at  [N/$\alpha$]$\sim -0.9$, although it could
be argued that there are in fact {\em two} plateaus,
the second found around [N/$\alpha$]$\sim -1.4$.

From the nucleosynthetic point of view
there is only one mode of nitrogen production:
it is formed in the course of the CNO cycle, 
at the expenses of available C and O.
Since this requires the pre-existence
of nuclei of C and O, one may expect the
production of N to increase with increasing
abundance of C and  O, i.e. with metallicity.
Nitrogen would then be a so-called purely
{\em secondary} element, to distinguish
it from {\em primary} elements, like the
$\alpha$-chain elements, whose yield is
independent of metallicity. The ratio of 
a purely {\em secondary} element to a primary one
should be a monotonically increasing function
of metallicity. The fact that 
[N/$\alpha$] shows a plateau at low metallicity
demonstrates that N cannot
be a purely secondary element.
We note here that although the concept
of {\em primary} and {\em secondary} elements is a useful
simplification, 
for the majority of the elements
the production history is complex enough that it is inappropriate
to label it with either of these terms.

For nitrogen, although the nucleosynthetic process is unique,
it is necessary to invoke different sites where the process
occurs to explain the two kinds of behaviour.
Intermediate mass
stars in the AGB phase
(IM-AGB stars, Ventura et al. 2002), in which the H-burning shell
operates through the CNO cycle, have always been considered
major producers of nitrogen.
For these stars the necessary C and O nuclei
are those already present at the time 
of the formation of the star and the production
mode would be {\em secondary}.
Massive stars are other possible N producers.  In 
such stars the central H burning 
takes place  through the CNO cycle, and again the
mode of production would be secondary. If, however,
the star is rotating, as soon as the central He 
burning is ignited, fresh C and O produced in the core
may be transported through rotational mixing to the H-burning
shell. This  would then produce N in a primary 
fashion, since the necessary parent nuclei are
created {\em in situ} (see Maeder and Meynet (2005) 
for a complete discussion).
In such a scenario N production cannot be
purely primary, since some secondary production
is unavoidable.
A primary mode of production of N may occur also
in IM-AGB stars, provided a way of mixing the 
products of the He burning shell to the
H burning shell exists.

The two above-mentioned
producers (IM-AGB stars and massive stars) release
their nucleosynthetic products, which become 
available for the next generation of stars, on 
quite different time-scales. Massive stars will
explode as core-collapse supernovae on time scales
shorter that 10 Myr, while IM-AGB will release their
products through winds or in the planetary nebula phase
on a time scale between 300 Myr and 1Gyr, depending on
their mass and metallicity.

The pattern of  N abundances in Globular Clusters is
somewhat more complex than in field stars.
In the first place, like other light elements,
N displays sizeable star to star scatter 
within the cluster. From studies conducted
with intermediate band photometry or
low resolution spectra it has been clear that 
many clusters display variations of N abundances 
\citep[see][for a review of these older results]{kraft}. 
For many years such effects were supposed to be 
caused by mixing in the giant stars, which were the only ones
it was possible to analyse 
\citep[see][for an interesting idea on the role of
rotation]{norris81}.
In the recent years, largely thanks to advent of
8m class telescopes, it has been possible
to investigate these effects down to the TO of
a few Globular Clusters, and the abundance variations
appear to persist. Since a TO star is not expected to
experience mixing, this calls for another
explanation for these variations
\citep{carretta05}.
Although it is generally accepted that the
stars we presently observe in GCs have been
formed out of material which has been
inhomogeneously polluted by a previous 
generation of stars, there is little
consensus on which stars were responsible
for the pollution.
Proposed candidates are
IM-AGB stars (D' Antona et al. 2005) and
massive stars (see e.g. Decressin et al. 2007a) , including Wolf-Rayet
objects (Smith 2006). 

It seems therefore timely to compare the N abundance observed in unevolved
field and Globular Cluster
stars, by making the  effort to place  them on the same abundance scale,
and  to use this comparison
to shine some light on the formation of nitrogen in the Galaxy and on
the formation process
 of Globular Clusters and on the Globulars - Halo relationship.

We here present abundances for five TO stars (3 \relax 
in NGC 6397
and 2 \relax in NGC 6752) derived from 
the compact band head of the
(0-0) vibrational band of the
NH $A^3\Pi-X^3\Sigma^-$ system
at 3360 \AA. This feature is stronger
than  the UV CN bands used in \citet{carretta05}
and should provide a more accurate measurement.
Moreover the N abundance derived from the NH bands
does not depend on either C nor O, instead when we
use the CN bands the derived N abundance depends on the
assumed C and O abundances. The O abundance enters
because the majority of C atoms is always locked
in the tightly bound CO molecule. 
For four out of the five stars of our sample
preliminary abundances based on the NH
bands have been presented in Pasquini et al. (2004, 2007),
however we deemed necessary  a reanalysis using
the same molecular data and analysis procedure
as in Ecuvillon et al. (2004) and Israelian et al. (2004),
to allow an unambiguous comparison of the N abundances
in the GCs with those in field stars of comparable
metallicity.

\section{Sample Stars and Observations}

Being the closest GCs in the southern hemisphere, NGC 6397 and NGC 6752
are amongst the most
studied GCs. 
Many high resolution spectra exist, also of TO  stars.

The objects and the observations presented here have been described in
Pasquini et al. (2004, 2007).
 Only star  1406 of NGC 6397  was not contained in the
Pasquini et al. (2004) sample, because the S/N ratio obtained in the Be
region was too low to
attempt any Be abundance analysis.

The spectra discussed in this paper are only the ones
taken with the blue arm of the UVES spectrograph (Dekker et al. 2000) 
on the ESO Kueyen-VLT 8.2m telescope. An identical setting
was used in the observations of the two Globular Clusters:
central wavelength 3460 \AA, 1\farcs{0} slit and $2\times2$
on-chip binning. The effective resolution was around
R$\sim 45\,000$. Further details can be found in 
in Pasquini et al. (2004, 2007).

The two stars of NGC 6752 were selected at
the extremes  of the O-Na anti-correlation
from Gratton et al. (2001):  star 4428 is
representative of the O-rich, Li rich component, while  star
200613 is representative of the O-poor - Li poor component. By observing
stars at the extremes of the
chemical distribution, we aimed at maximising the chances of observing
possible differences
in N abundance.

Table 1 summarises the characteristics of the stars, including the
abundances of the single elements, as derived from the literature. The
[N/H] as derived in  the
present work is also given, in the  last column. The stellar parameters
listed in Table 1 are those adopted in our spectroscopic analysis. Only
abundances of those
elements which are known to vary from star to star (O, Na, Li, N) are
listed in Table 1.
The reader can find  additional element abundances
in the quoted papers: G01, James et al. (2004), and Carretta et
al. (2005).

\begin{table*}
\caption{NGC~6752 sample stars, their atmospheric parameters and
abundances. The atmospheric parameters,  [Fe/H], [O/H] and [Na/Fe] are
from Gratton et al. 2001;  Log(Li/H) from Pasquini et al. (2005),
and [Be/H] from Pasquini et al. 2006, [N/H] from this work }

\begin{tabular}{lllllllll}
\hline
Star  & T$_{\rm eff}$ & $\log g$   & [Fe/H] & [O/H]   & [Na/Fe]
&  $\log({\rm Li/H})$  & $\log({\rm Be/H})$  & [N/H] \\
            &           &           &           &     &            &
                       &       &         \\
4428       & 6226      &  4.28     &  $-1.52$ & $-1.28$   &  $-0.35  $
& 2.50      &   $-12.04  $      &  $-1.2 \pm 0.1$ \\
200613   & 6226      &  4.28     &  $-1.56$ & $ / $         &  $0.64 $
& 2.13      &   $< -12.2 $     &  0 $\pm 0.1$   \\
\hline
\end{tabular}
\end{table*}

\begin{table*}
\caption{NGC~6397 stars, their atmospheric parameters and abundances. The
atmospheric parameters, are  from Bonifacio  et al. (2002), 
[Be/H] from Pasquini et al. 2004, [N/H] from this work }

\begin{tabular}{lllllllll}
\hline
Star  & T$_{\rm eff}$ & $\log g$   & [Fe/H] & [O/H]   & [Na/Fe]
&  $\log({\rm Li/H})$ &  $\log({\rm Be/H})$  & [N/H]  \\
         &           &           &           &     &             &      &
                &         \\
2111     &  6207     &  4.1    &  $-2.01 $ & $-2.24$    &         +0.17
&  2.33      &   $-12.27$      &  $-0.76 \pm 0.1$     \\
228       &  6274     &  4.1    &  $-2.05 $  & $-1.64$   &          +0.16
&  2.28      &    $ -12.43$    &   $-0.75  \pm 0.15$  \\
1406     &  6345     & 4.1      &  $ -2.04 $ &        /         &    / & 
  2.37          &       /              & $ -1 \pm 0.2 $
\\ \hline
\end{tabular}
\end{table*}

\section{Abundance analysis}

The main aim of this paper is to set
the cluster stars on the same
N abundance scale  as  
field stars (Israelian et al. 2004, Ecuvillon et al. 2004).
We therefore used exactly the same line fitting method, 
molecular and atomic line data as Ecuvillon et al. (2004).
The model atmospheres where the same used in 
Pasquini et al. (2004, 2007)
and were computed with version 9 of the ATLAS
code  
\citep{1993KurCD..13.....K,2005MSAIS...8...14K}
its Linux version \citep{2004MSAIS...5...93S,2005MSAIS...8...61S}.
All the models were computed
with the
``NEW'' Opacity Distribution Functions 
\citep{2003IAUS..210P.A20C}, 
with 1\kms\ micro-turbulence,  a
mixing-length parameter~\mlp\ of 1.25
and no overshooting.
The synthetic spectra were computed
with the
line formation code
MOOG\citep{sneden73,sneden74,snedenweb}.

\begin{figure}[]
\begin{center}
 \resizebox{\hsize}{!}{\includegraphics{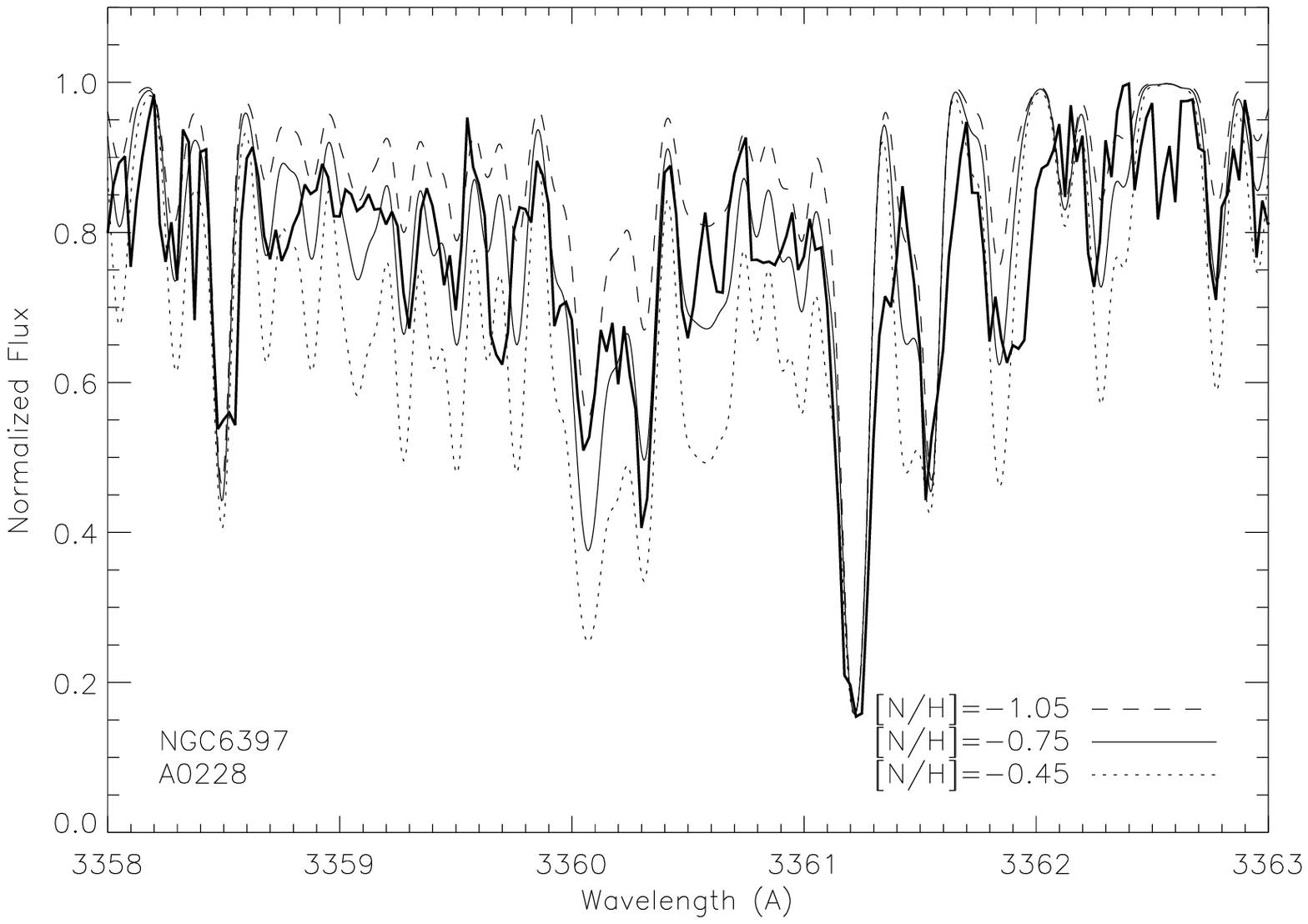}}
 \resizebox{\hsize}{!}{\includegraphics{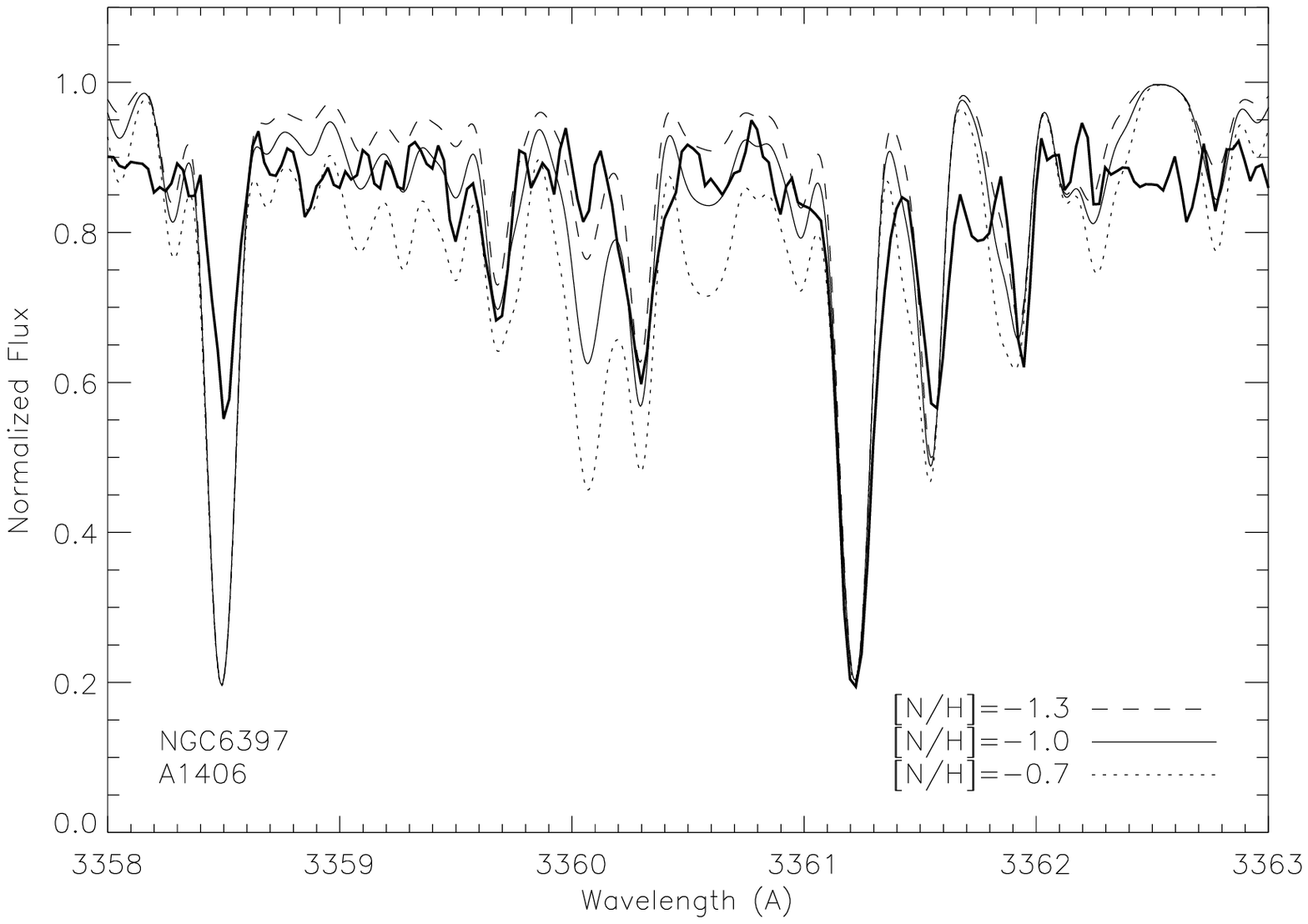}}
 \resizebox{\hsize}{!}{\includegraphics{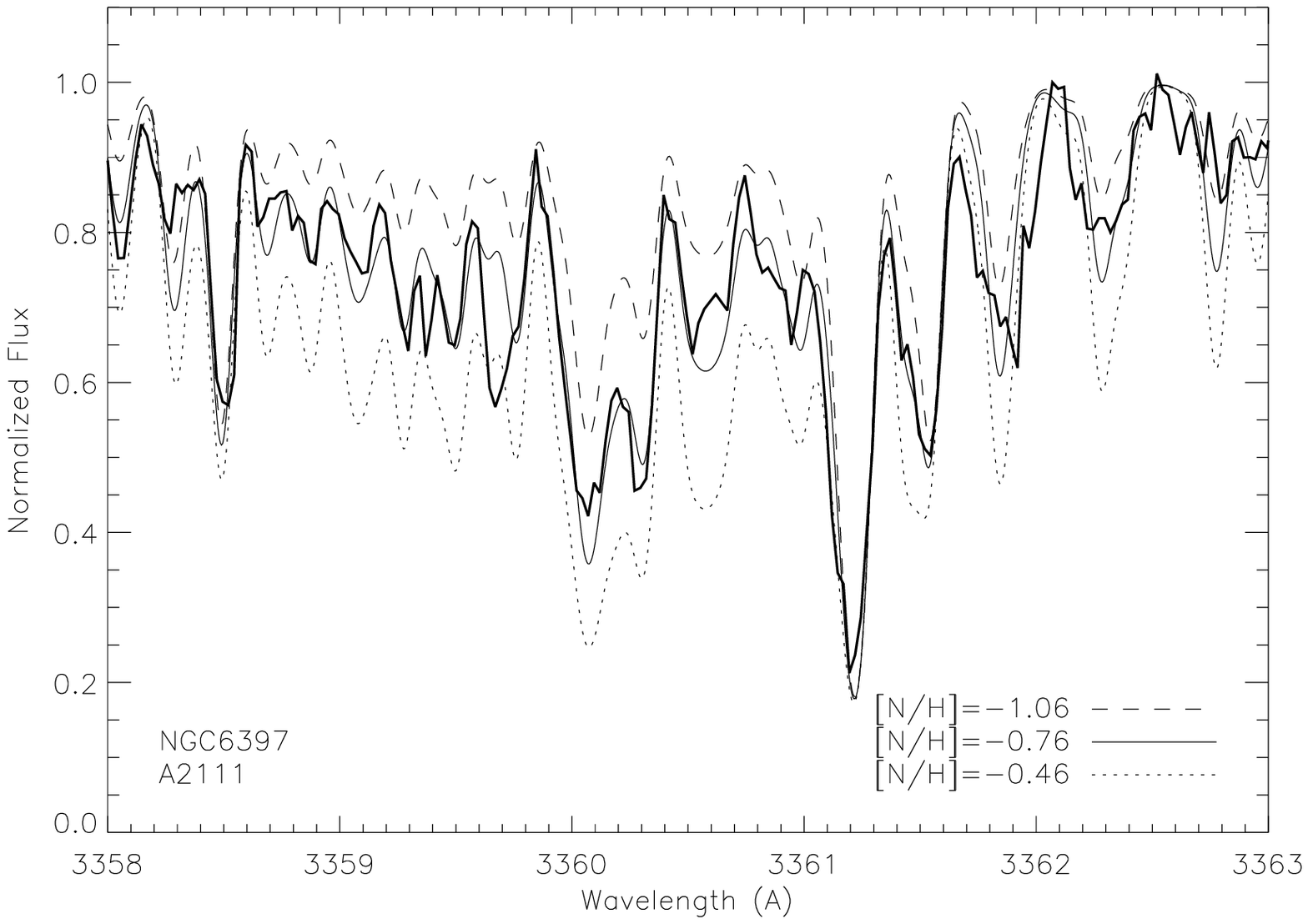}}
\end{center}
\caption{\label{obs6397}UVES spectra of the NGC 6397  TO stars in the NH band.
best fit models are shown in bold, while models varying by $\pm$0.3 dex  in NH are
shown as dashed and dotted lines.}
\end{figure}

\begin{figure}[]
\begin{center}
 \resizebox{\hsize}{!}{\includegraphics{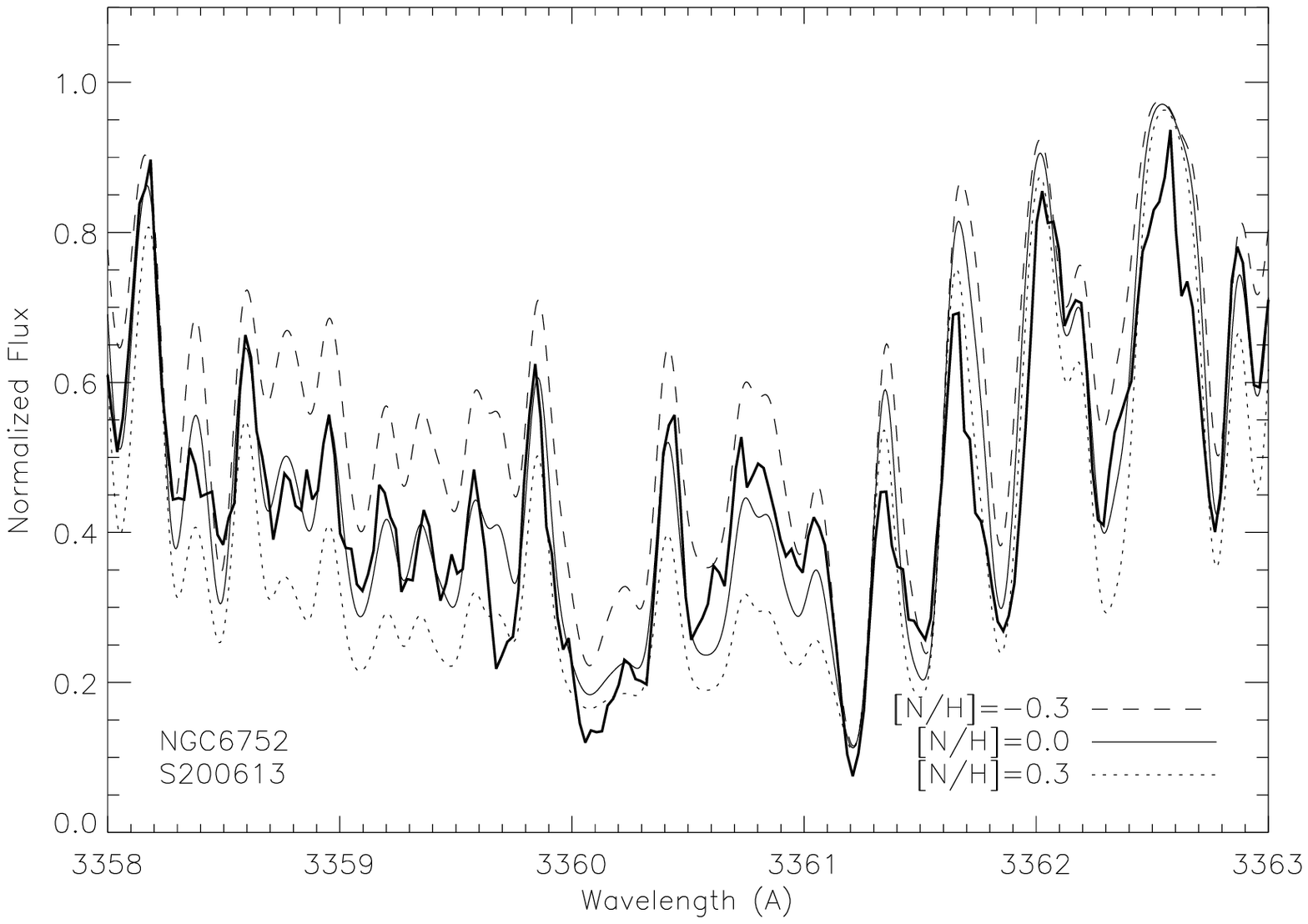}}
 \resizebox{\hsize}{!}{\includegraphics{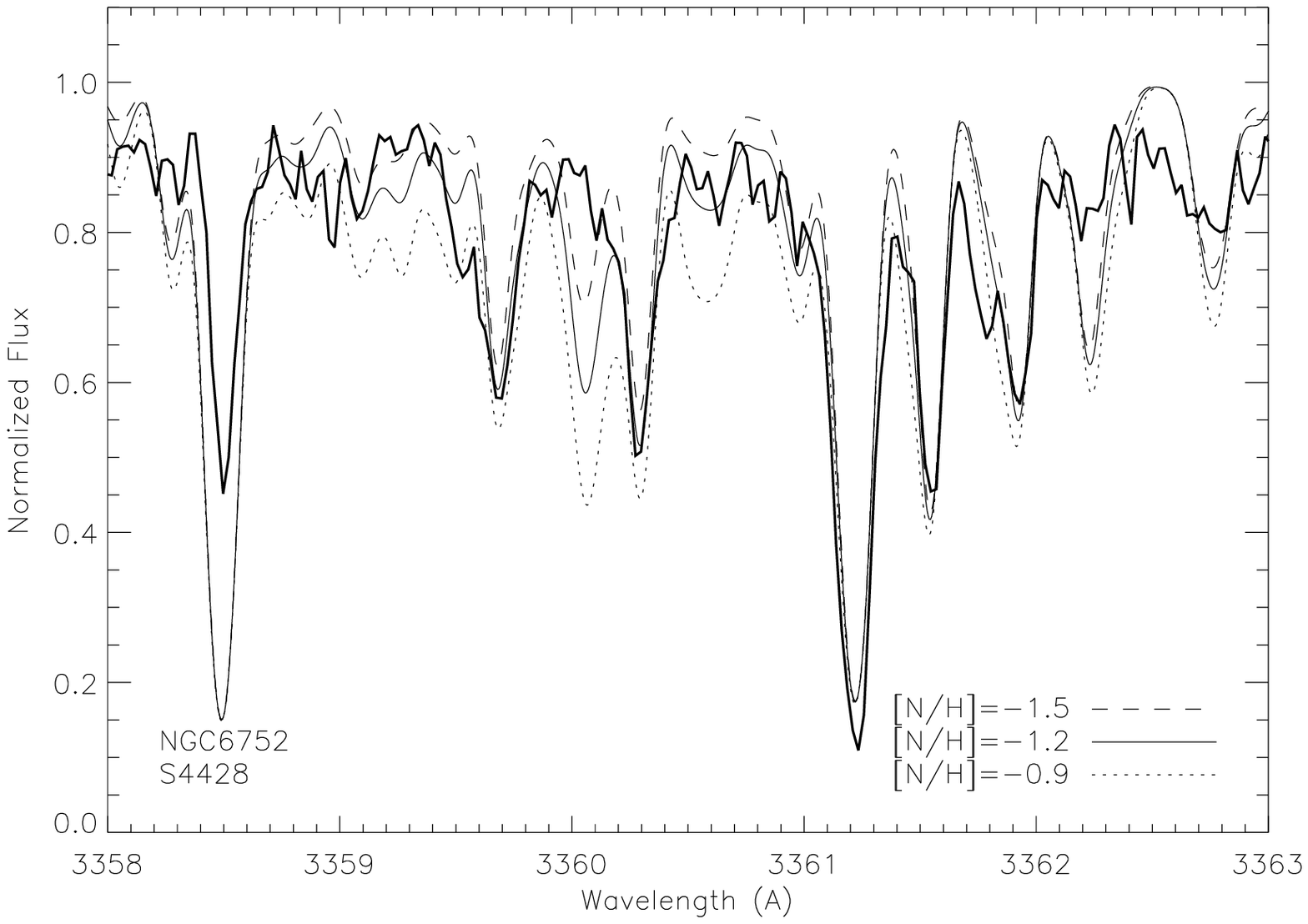}}
\end{center}
\caption{\label{obs6752}UVES spectra of the NGC 6752 TO stars in the NH band.
best fit models are shown in bold, while models varying by $\pm$0.3 dex in NH are
shown as dashed and dotted lines.}
\end{figure}

Figures \ref{obs6397} and \ref{obs6752}  
show the spectra of the five stars  compared to
synthetic spectra. The synthetic spectrum with the 
best fit 
N abundance is always shown as a solid line.

The stellar parameters adopted  are given in Table 1. 
A difference in  T$_{eff}$
of $\pm$100 K implies a change in  [N/H] of $\pm 0.09$ dex. 
In spite of the differences in atomic and molecular data involved,
and of the line formation code used, 
the comparison between the present results and those of Pasquini et
al. (2004, 2007) is excellent (agreement to better than 0.05 dex),
except for star 4428 of NGC 6752, for which there is a discrepancy of 
a factor of two.
In spite of this discrepancy we 
confirm the huge abundance difference (1.3 dex)  between the two
observed stars of this cluster.

 Carretta et al. (2005) used in their analysis the CN band at 3880 \AA\ to
 derive N in several stars in GCs.
 For all the dwarfs in NGC 6397 they found only upper limits, higher
 (therefore consistent) than our measurements.
 As far as the NGC 6752 dwarfs are concerned, they quote a similar value
 to ours for star 200613, 
the value 10 times higher quoted for star 4428 in that paper, should instead
be considered as an upper limit (E. Carretta 
private communication, see also Pasquini et al. 2007).

\begin{figure}[]
\begin{center}
\resizebox{\hsize}{!}{\includegraphics{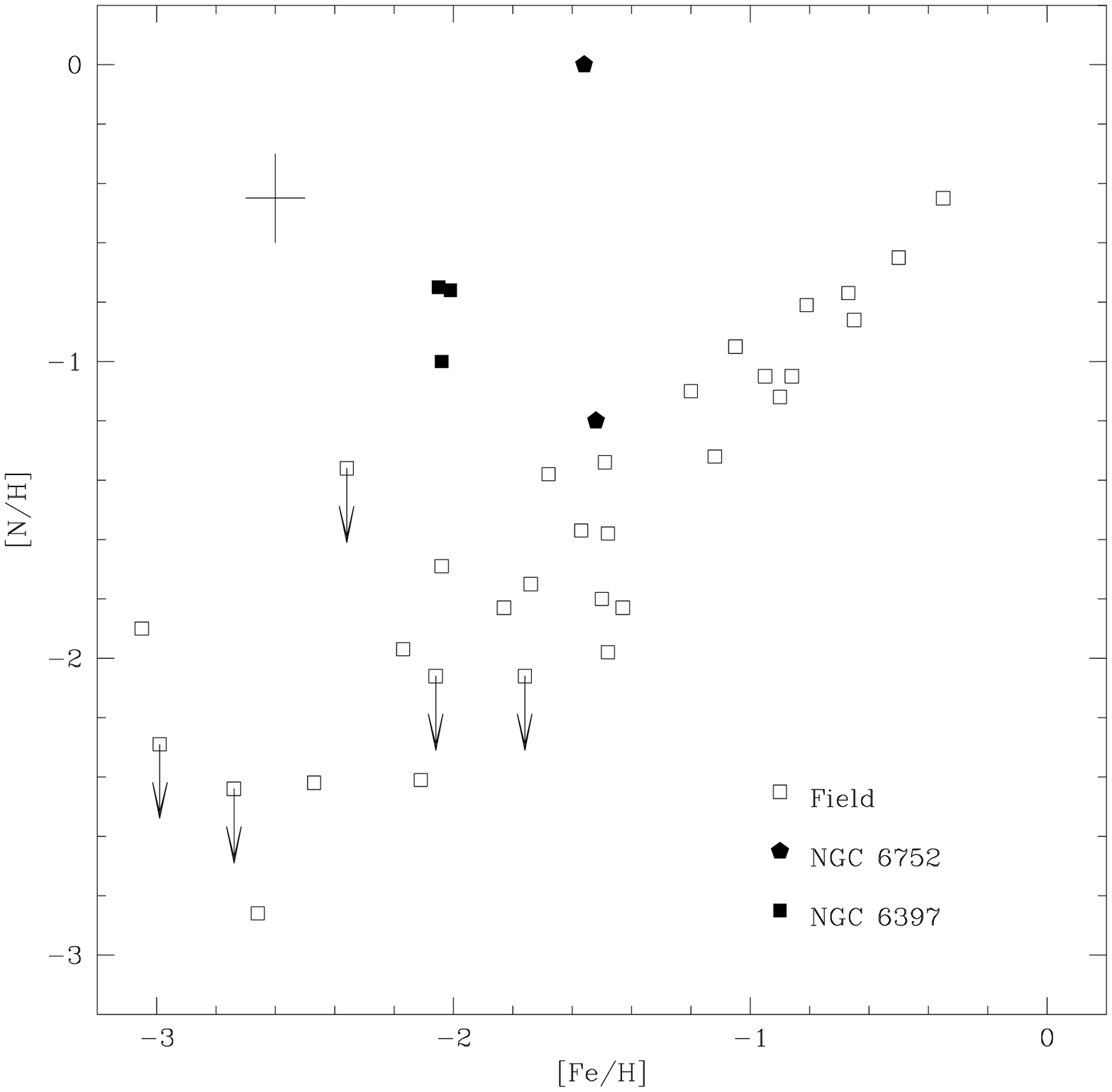}}
\end{center}
\caption{ \label{fignfe}[N/H] vs. [Fe/H] for the field stars of Israelian et al. (2004)
and the
TO stars of the two globular clusters NGC 6397 and NGC 6752  (this
paper). The typical
error bar is given on the top left corner.}
\end{figure}

\begin{figure}[]
\begin{center}
\resizebox{\hsize}{!}{\includegraphics{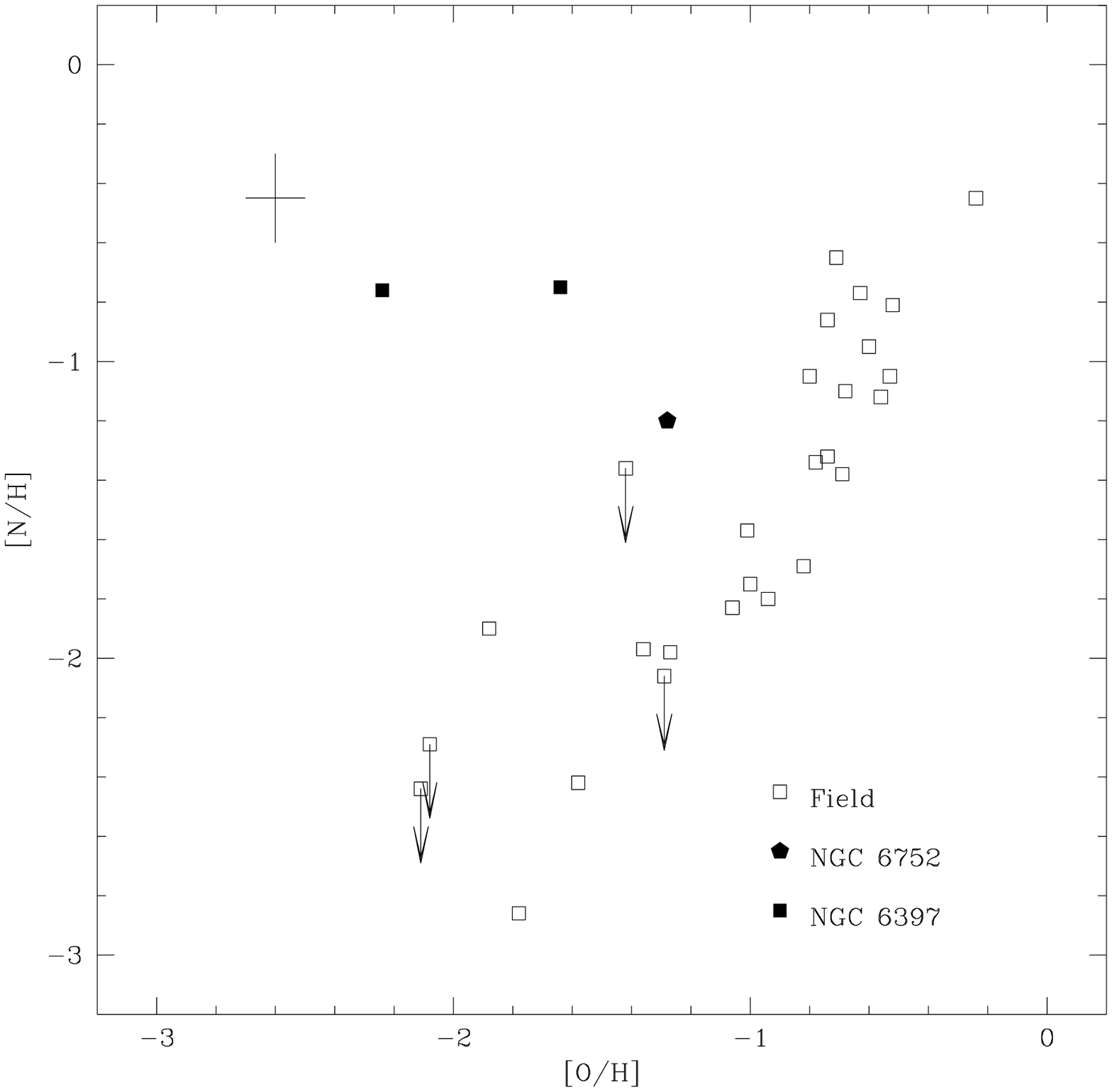}}
\end{center}
\caption{ \label{figno}[N/H] vs. [O/H] for the field stars of Israelian et al. (2004)
and the
TO stars of the two globular clusters NGC 6397 and NGC 6752  (this
paper). The typical
error bar is given on the top left corner.}
\end{figure}

\section{Discussion}

\subsection{ Comparison with  field stars}

Our N abundances for  GC stars can be
consistently compared to the results for field stars
of Ecuvillon et al. (2004) and Israelian et al. (2004).

Figure \ref{fignfe} shows the [N/H] abundance  vs. [Fe/H] for field and cluster
stars. Most field stars closely follow
a 1:1 relationship, with a scatter compatible with the error measurements
(quoted as $\sim \pm$ 0.15
by Israelian et al. 2004 and given as a reference in the upper left
corner of the Figure).
The [N/O] ratio
is shown in Fig. \ref{figno}, for
the present discussion we ignore 
the possible complications arising from the fact 
that the O abundances displayed have been 
derived in an inhomogeneous way and using different 
O indicators, and take all measurements at face
value. With the {\em caveat} that any systematic differences
between the different analysis will tend to 
blur any existing pattern.
Two points  emerge  from Figure 2 :

\begin{enumerate}

\item  The cluster stars do not resemble, in general, the behaviour of
the field stars.

\item  The two clusters do  not show the  same behaviour.

\end{enumerate}

The first point is somewhat expected, since we 
have known  for long time of the
presence of chemical
anomalies in GC stars, which includes the presence of N rich stars. N rich stars,
though known in the  field,   are extremely rare, and a
typical sample, like
the one of Israelian et al. (2004), contains at most one  such object.

The two  NGC 6752 stars show a large spread in N (almost
a factor of 20) and
the cluster stars, including all stars  observed in  NGC 6397, have a
very high N abundance,
clearly  departing from the field stars relationship at high significance.
This is clearly related to the peculiar formation mechanism of GCs,
and we will
come back on this point in the next section.

As far as the second point is concerned  
we believe  that this aspect should be
considered with some attention, since it has not been explored too
much so far.
For this reason we look at the two clusters separately.

NGC 6752 is a cluster showing all possible chemical inhomogeneities
and anticorrelations.
But, in its anomalies,  the observed trends are  extremely consistent.
The difference of a factor $\sim$20 in N among the two stars is very
high, but
it is in line with the other large differences recorded for them in Na
(factor 10)  and  Li
(factor 2.5 ) in these stars
and in oxygen (factor 4) in other stars having
abundance pattern  similar to star 200613
(in 200613 O has not been
measured).

With our new value, the N of star 4428 is  higher, but formally
compatible with
the field stars having similar metallicity.
Even if we do not
know how this GC formed, the observations are  compatible with the
idea that star 4428 is the prototype of an 'unpolluted' star, perfectly
similar in all aspects to
its field counterparts of similar [Fe/H],
while star 200613 is, at the opposite, the prototype of  highly polluted
star, with high N, high Na, low Li and low O, showing
therefore the signature
of gas which experienced high temperatures where 
nuclear cycles such as CNO,
Ne-Na, and
the consequent Li destruction, have occurred.

When plotting N vs. O (Fig. \ref{figno}),  the situation is similar, 
unfortunately we have O abundances only for 3 out of five stars.
The impression is 
that the cluster stars
(including star 4428)  stand definitely out of the  field relationship.
This may be related to the fact that star 4428 has an O/Fe ratio
which is lower, by roughly a factor of 2, than field stars of
the same [Fe/H], when N is plotted against [Fe/H] star 4428
follows the trend of field stars, but when plotted against
oxygen it appears to have too much nitrogen, with respect
to field stars with the same [O/H]. Clearly in a cluster
like NGC 6752, which has a very strong Na-O anticorrelation,
it is difficult to tell which is the ``unpolluted'' oxygen
content of the cluster. Among the stars analysed by Gratton et al.
(2001) there is one dwarf and two subgiants which have higher
[O/H] than star 4428. We have chosen star 4428 as template
of the ``unpolluted'' population, because it is the star
with the lowest [Na/Fe]. New measurements of N, based on the
NH bands for all the stars observed by Gratton et al. (2001),
would be desirable, given the weakness of the CN bands 
and the difficulty to use them to measure N.

Summarising, for NGC 6752, our results confirm that the chemical
composition of the stars
is consistent with the  assumption that each object is composed by a
mixture of  'pristine'
material, possibly N rich, but otherwise indistinguishable  from that
forming typical halo stars,
with highly processed material, N and Na rich, O and  Li  poor.

The NGC 6397 cluster stars are on the other hand more intriguing:
the NGC 6397 objects have  very similar N abundance, with a negligible
spread,  but  at an absolute level about 10 times higher than field
stars of similar
metallicity.
It is not possible, even taking the highest rate of rapid rotating
massive stars, to insert such a high N production into a standard Galactic chemical evolutionary
model (see e.g. Chiappini et al. 2006), thus such a high N is the direct proof that the NGC 6397 
stars have been affected by  a peculiar  process. 
In their extensive study of the Globular Cluster system
of M 31 \citet{burstein}, using integrated light spectra, 
have shown that the older clusters of M 31 have considerably stronger
NH bands than Galactic clusters.
The reasons for this difference is unclear. 
NGC 6397 is not among the Galactic Globular Clusters considered
by   \citet{burstein}, but adopting the naive point of view that
NH band strength is related to N abundance, one may expect
that NGC 6397 \relax in integrated light should show a stronger
NH band than other   Galactic Globular Clusters, although
not necessarily as strong as that of the M 31 clusters.
It remains to be investigated if the mechanism which causes a
strong NH band in the Clusters of M 31 is the same, or akin to,
that which causes the high N abundance in NGC 6397.

The oxygen content of the cluster is, on the opposite,  very  low;
at least a factor of two lower than
the field stars of similar metallicity.
Even if  the behaviour of oxygen in metal poor stars is highly debated,
oxygen has been found at the level of only [O/Fe] = 0.2-0.3 in NGC 6397
by all authors
(Th\'evenin et al. 2001, Pasquini et al. 2004,  Gratton et al. 2001).
Carretta et al. 2005 show that the C in NGC~6397   is at a level of
[C/Fe]$\sim$0,  which is very similar to what is observed in field stars
of similar metallicity.
This  clearly indicates that for the dwarfs observed in this  cluster
the whole CNO balance
is different to that of field stars.
The Li abundance 
(Molaro \& Pasquini, 1994, Pasquini \& Molaro 1996,
 Th\'evenin et al. 2001, Bonifacio et al. 2002, Korn et al. 2006)
is, on the other hand, 
absolutely consistent with the {\em Spite plateau} \citep{spite82a,spite82b},  
and  Na, with a slightly supersolar value,
does not vary at a significant level in the stars so far observed (Gratton et al. 2001).

As far as the chemical composition of the TO stars is concerned, the
homogeneity in NGC 6397 stars seems much higher than in NGC 6752, in
that  only a large difference of oxygen has been reported by Pasquini
et al. (2004) for  two stars
of this sample, while no substantial spread is observed in any other
element among the main sequence stars.
Only three subgiants  show evidence for strong Na and N variations and
no real 'anticorrelation'
can be deduced by the abundances so far published (Gratton et al. 2001, 
Carretta et
al. 2005, Table 1).

Of course we must be careful in deriving strong conclusions from such
a low statistics,
and we are aware that some of the previous evidences (or missing
evidences) could be simply
the result of a combination of a limited sample and large errors, however,
we think we may summarise this part of the discussion saying that, in
spite of some similarity,  the two clusters show a different behaviour
with respect to their N content and, in general,
the spread of  light elements amongst their stars:
while the NGC 6752 stars abundances show consistent signatures of
pollution,
the NGC 6397 stars show at the same time the signatures of  CNO (and
likely NeNa)
processed  material, while showing at the same time no clear
anticorrelation,
a much larger homogeneity in  abundances, and a Li content similar to the
{\em Spite plateau}.

\subsection{About GC formation}

One fundamental consideration in the discussion of the formation
scenarios is the fact that  anti-correlations persist in evolved stars after the
dredge-up (see e.g. Grundhal et al. 2002).
This shows that the whole stellar mass, and  not only the external
stellar layers, is affected by the chemical anomalies.

 As far as this paper is concerned, we may summarise the on-going
 discussion on the formation of GC in two
 main points:

 \begin{itemize}
 \item The mass of the ``polluting'' stars:  broadly speaking some groups
 advocate that the CNO cycling has been
 produced by intermediate mass AGB stars(Ventura et al. 2002, D' Antona
et al. 2005), while others
 favour  massive rotating stars (Maeder and Meynet 2002, Decressin et
 al. 2007b).
 Each of these schemes has strong and weak points. In an attempt to
 summarise them,  we can say that  IM-AGB stars have the advantage of
 being 'natural' N  producers,
 able to also produce Li (and explain the behaviour of NGC 6397 through the
 so called Cameron-Fowler  mechanism; Cameron \& Fowler, 1971), 
but having at present two
 main problems: the
 difficulty of reproducing the  observed yields for other elements
 (in particular Na), the fact that
 for NGC 6397 a natural conspiracy should have produced an ad hoc amount
 of Li,  exactly the same
 observed in the Spite plateau (Bonifacio et al. 2002). Finally,  the
 IM-AGB hypothesis would require a very peculiar IMF,
 heavily flat-topped (Prantzos and Charbonnel 2005, Chiappini et
 al. 2006).
 As far the rotating massive hypothesis is concerned,
 this scheme mitigates the IMF requirements, and can better explain some
 of the observed yields,
 although even the most extreme models cannot account for the observed
 high range of the Mg-Al
 correlation, unless
 some of the employed
nuclear cross-sections are affected by large errors 
(Decressin et al. 2007b).

 One important conclusion of both hypotheses is that the ``polluted'' 
 stars should have
 a much higher  $^4$He abundance. Higher $^4$He abundance is claimed to
 be observed in one of the Main Sequences of $\Omega$ Cen \citep{piotto05}
and
of NGC 2808 \citep{piotto2808};
 it would naturally explain the blue extension of
 HB in NGC 6752 and other clusters (D'Antona et al. 2005). On the other
 hand, a substantially higher $^4$He abundance, should
 spectroscopically produce  an apparent higher abundance of heavier
 elements and higher
 spectroscopic gravity  in $^4$He rich stars (B\"ohm-Vitense 1979).
A spread in He abundances, should bring about
a spread in the abundances of {\em all} elements, including 
iron-peak elements. The results of Gratton et al. (2001) 
in NGC 6397 and NGC 6752 exclude any significant spread 
in Fe abundances in either cluster. 

 \item The second open issue is related to the overall mass of the
 structures containing the contaminants.
 Most analysis have so-far analysed self-pollution, that is,  that
 all the
 polluted material was   created within the cluster.
 New schemes consider the possibility of infall from
 the exterior (e.g. Bekki et al. 2007). In this scheme, for instance,
 GCs are
 the remnants of the core of much more massive dwarf  galaxies,  receiving
 processed gas
 from the whole galaxy, rather than from its own stars only. This
 scheme would
 greatly mitigate the quest for anomalous IMF, but , on the other hand, it
 opens the parameter space , since a new variable (the infall of material)
 is now added, and  it has not been  looked in detail as far as chemical
 evolution is concerned.

 We shall finally recall that most attempts  aim to explain
 all the data observed in the different GC within a unique scheme.
 This is , of course, a very ambitious and valuable approach, but we
 would like to
 point out that the observational data 
show that different clusters present a
 different behaviour, and
 therefore, the attempt to unify all observations in a unique scheme
 might be premature.

 \end{itemize}

The measured abundances give us the possibility of studying the 
composition of  the polluting gas in some detail.  
For NGC 6752 we can safely assume that the two stars observed are at
the extremes of the chemical distribution of the stars in the cluster.

In order to analyse pollution, Li is the best element to be considered,
since it is destroyed at the temperatures where CNO cycling occurs. The difference
of Li between the two stars is of a factor about 2.5.  If no Li production has occurred,
 this implies a fraction of  60$\%$ of processed material in the
 contaminated star.
 The observed N enhancement would  imply a 25-fold
   N production in the polluting  gas. This can be assumed as a solid
   lower limit for the yields, in case no infall is present.
 In case the polluting gas is somewhat rich in Li the amount of 
   polluting gas  would be larger than  60$\%$ and, as a consequence, the
    overproduction of N in the previous generation  lower than what estimated above. 
  We note that a 25-fold  N production is
      already at the upper limits  for  the current models of   massive stars.
  Detailed production predictions have been computed e.g. by Decressin et al. (2007b), 
   who find that  massive stars can produce in their winds up to [N/Fe]$\sim$ 1.5, or a factor 
    30 at most.  
  
  The constraints from oxygen are also relevant: the ratio in O abundance
   between star 4428 (O-rich) and the most O-poor stars observed in
    NGC 6752 is  of about 4.5 (Carretta et al. 2005), which would put the lower 
     limit constraint of 22.5$\%$ to the amount of pristine material,
      in case the processed material had no O at all.

The NGC 6752 data are consistent with a picture where the stars in 
this cluster   are formed by a mixture   of ``pristine'' and ``processed'' 
material.  The maximum observed
``pollution''   amounts 
to about 60 $\%$ of the gas forming the stars. This would
imply a polluting material  with no  Li, no  O, and 
N enhanced by 25 times (and Na  by a  similar amount).
It may be difficult to produce in a closed system
the very   high amount of N required by this scenario.
The oxygen difference 
among the extreme stars suggests an even higher level of
polluted material, which would require  some Li production, 
and  in this case the 
N (and Na) overabundance should be proportionally lower in the
polluting gas.  
 
The abundances of NGC 6397   cannot be accounted by a simple 
 pollution picture, since  the basic variations in oxygen are
not accompanied by
similar variations in Li  (Bonifacio et al. 2002) and   in N (this paper). 

\begin{acknowledgements}
P.B. acknowledges financial
support from EU contract MEXT-CT-2004-014265 (CIFIST).
\end{acknowledgements}
\bibliographystyle{aa}

\end{document}